\documentclass[conference]{IEEEtran}
\IEEEoverridecommandlockouts
\usepackage{cite}
\usepackage{amsmath,amssymb,amsfonts}
\usepackage{algorithmic}
\usepackage{graphicx}
\usepackage{textcomp}
\usepackage{xcolor}
\usepackage{lipsum,comment}
\usepackage{array}

\begin{document}
\title{Mixer: DNN Watermarking using Image Mixup}

\author{
\IEEEauthorblockN{Kassem Kallas and Teddy Furon}
\IEEEauthorblockA{Centre Inria de l'Université de Rennes, France \\
firstname.lastname@inria.fr}
}

\maketitle

\begin{abstract}\label{abstract}
It is crucial to protect the intellectual property rights of DNN models prior to their deployment.
The DNN should perform two main tasks: its primary task and watermarking task. This paper proposes a lightweight, reliable, and secure DNN watermarking that attempts to establish strong ties between these two tasks. The samples triggering the watermarking task are generated using image Mixup either from training or testing samples. This means that there is an infinity of triggers not limited to the samples used to embed the watermark in the model at training. 
The extensive experiments on image classification models for different datasets as well as exposing them to a variety of attacks, show that the proposed watermarking provides protection with an adequate level of security and robustness.
\end{abstract}


\section{Introduction}\label{sec:introduction}
Deep Neural Networks (DNNs) became the de facto standards for a plethora of computer vision tasks, including image classification. It is imperative to protect DNN ownership because they are regarded as valuable industrial assets~\cite{SurveyDeployingDNN2022}.
The literature demonstrates that it is technically feasible to watermark DNNs without spoiling its accuracy. Removing the watermark from a model that happens to be equally difficult as training a new network from scratch. This justifies that DNN watermarking is a primary tool for ownership protection.

A popular watermarking of classification models is that the Owner injects few inputs with unrelated labels in the training set so that the model keeps a memory of them. Once deployed, the Verifier submits these samples (a.k.a. triggers) and checks whether their outputs corresponds to the unrelated labels. A non-watermarked model would not classify these triggers in the same surprising way. The first difficulty is that this watermarking modifies very locally the behavior of the model which might be partially erased if the Attacker transforms the model. Another point is that the triggers play the role of the secret watermarking key. Once disclosed, the Attacker can remove the watermark by making the model forgetting about these isolated inputs.    

Our paper focuses on the latter threat. Our goal is to enforce that the samples used to embed the watermark at training time are different than the samples used for detecting the watermark at verification time. Indeed, the secret key of our watermarking scheme is no longer a small set of isolated inputs but a manifold of inputs. There is thus an infinite number of triggering samples and the Verifier never uses twice the same samples.     
According to experimental analysis on image classification with various datasets and DNN architectures, our watermarking does not reduce the accuracy of the model while being robust to classic post-processing. Indeed, strong attacks remove the watermark at some point but also ruin the classification accuracy. 

\section{Prior Art}\label{sec:background}
The first DNN watermaking occurred in 2017, as per Uchida et al. in~\cite{uchida2017}. It is a white box setting in the sense that the watermark is embedded (detected) directly into (resp. from) the host DNN parameters. This approach was improved in~\cite{SpreadDitherWatermark2021}.
In the black box setting, the watermark modifies the behavior of the host DNN and this can be testify just by querying inputs and observing outputs at verification time. Backdooring is such an example by forcing the DNN to overfit over a unusual set of image-label pairs and thus, making them difficult to be predicted by any other model. These pairs can be referred to as, backdoors, triggers, watermarked samples or secret keys. They can be synthetic or unnatural~\cite{backdoorWM2018}, adversarial~\cite{Adversarialfrontier2020}, or benign samples with either visible~\cite{ProtectingIntellectualProperty2018} or invisible~\cite{Backdoorsignatureembeddedsystems2018,NoiseKeyImagesPaper2020} overlay.

As for the attacks, the three most widely used post-processing techniques for evaluating the robustness of DNN watermarking are transfer learning, model pruning, and weight quantization. Pruning and weight quantization are common processing for lowering the memory footprint of the network.
Transfer learning typically involves replacing the fully connected layers of a DNN with new layers adequate for the task the knowledge is transfered to i.e. use a pretrained DNN on ImageNet~\cite{Imagenet2009} to classify CIFAR10~\cite{cifar102009}. A particular case is fine-tuning in which a trained DNN is re-trained with a lower learning rate.
An attack is regarded as successful if and only if it eliminates the watermark behavior while maintaining the model accuracy on the main task descent. Otherwise, the attacker ends up receiving a needless DNN model.

More powerful attacks exist. 
Paper~\cite{ExponentialWeighting} demonstrates that reforming the inputs (with an auto-encoder for instance) removes the invisible overlay or the adversarial perturbation because these signals are too fragile.
The authors of~\cite{FourchallengesandFuneral2021} make the analogy with classic media watermarking where security is defined as the inability for the attacker to estimate the secret key.
For instance, a visible overlay can be easily copied once disclosed, which opens the door to ownership usurpation.
Using natural images with random labels as triggers prevents reforming or overlay copy attack. For instance, paper~\cite{kallas2022rose} imposes a specific generation of the random labels so that no Usurper can claim any existing model.   
Yet, these specific natural inputs constitute the secret key which is used at the embedding but also at the verification. Remember that a small fine-tuning easily makes the model forget about them if disclosed.
As far as we know, only paper~\cite{ProtocolVerification} protects these secret inputs by drowning them into a large pool of queried images at verification time to prevent their identification.

Our paper tackles this last pitfall because its secret key is not a small finite set of trigger images but an infinite trigger manifold. Therefore the triggers used at the verification are always different than the ones used at the embedding. In a way, this is equivalent of asymmetric media watermarking proposed in~\cite{furon1999asymmetric,furon2003asymmetric,boato2006improved}.    
\def\real{\mathbb{R}}
\def\C{\mathcal{C}}
\def\DT{\mathcal{D}_{train}}
\def\DTs{\mathcal{D}_{test}}
\def\Pr{\mathbb{P}}

\section{Proposed Method}\label{sec:proposedmethod}
\subsection{Image Mixup}
Mixup is a data-agnostic data augmentation technique used to construct new virtual training examples. Mixup is first introduced in~\cite{mixup2017} and extended to aligned mixup in~\cite{Shashanka_Mixup_2022}.  For instance, a new virtual example $\tilde{x}$ is constructed from two inputs $x_i$ and $x_j$ that could belongs to the same or different classes, as follows:
\begin{equation}
\label{eq:MixupImages}
    \tilde{x} = \lambda x_i + (1-\lambda) x_j, \quad 0<\lambda < 1.
\end{equation}
Mixup vicinal distribution is a type of regularization that promotes the model to act linearly. This lowers the number of unfavorable oscillations that occur when predicting outside of training samples.

\subsection{Mixup based watermark embedding}
\def\lambdab{\boldsymbol{\lambda}}
\def\alphab{\boldsymbol{\alpha}}
\def\mub{\boldsymbol{\mu}}
First, the proposed method extends the idea of mixup to use two or more inputs to generate the trigger samples, and moreover these inputs are random, denoted by capital letters:
\begin{equation}
\label{eq:OurMixupImages}
    \tilde{X} = \mathsf{clip}\left(\sum_{i=1}^{C} \lambda_i X_i + x_o\right)
\end{equation}
where image $X_i$ is one random input of class $i\in\{1,\ldots,C\}$, $\lambda_i>0$ is the weight assigned to the $i$-{th} class, and $x_o$ a visible constant additive overlay. As for the associated class, we also resort to a mixup with weighting vector $\mub$:
\begin{equation}
    \mathbf{\tilde{y}} = \sum_{i=1}^{C} \mu_i \mathbf{y}_i,
\end{equation}
where $\mathbf{y}_i$ is the one-hot vector corresponding to the $i$-th class.

The weighting vectors $\lambdab$ and $\mub$ are the secret key. They both are randomly drawn from the key distribution which is a compound of Dirichlet distributions. We first draw a random vector from the Dirichlet distribution of parameter $\alphab= (\alpha_1,\ldots,\alpha_C)$, and then we apply a random permutation to get $\lambdab$. This means that $\lambda_i\geq0$ and $\sum_i \lambda_i=1$. We repeat this process independently to get $\mub$. The profile $\alphab$ may contain some null values which implies that some classes receive a null weight. We denote by $m$ the number of non zero values in $\alphab$ which is in turn the number of non zero values in $\lambdab$ and $\mub$.

The owner randomly generates a set $S_{e}$ of $n_e$ mixup images associated with probability vector $\mathbf{\tilde{y}}$ and inject them in the training set.

\subsection{Mixup based watermark detection}
The Owner gives the keys $(\lambdab,\mub, x_o)$ to the Verifier who crafts a set $S_d$ of $n_d$ mixup images queried to the black box model.
Importantly, the set $S_d$ is different than the set $S_e$: The model $m$ under scrutiny is queried images not seen at training.
The Verifier computes the ratio
\begin{equation}
\label{eq:rho}
    \rho_{n_d} = |\{x\in S_d| \arg\max \mu_i m(x)_i = \arg\max \mu_i\}|/n_d,
\end{equation}
and decides that the model in the black box is watermarked if $\rho_{n_d}$ is bigger than a threshold $\tau$ which grants its ownership to the claimed Owner. 

\subsection{Analysis}
There are two hypotheses: the black box is or is not watermarked with the key $(\lambdab,\mub,x_o)$.
The Verifier crafts and queries $n_d$ new mixup samples and compute $\rho_{n_d}$. This is an empirical probability measured over $n_d$ trials. We denote by $\rho_P$ (resp. $\rho_N$) the true probability in the positive (resp. negative) case when the black box is (resp. is not) $(\lambdab,\mub,x_o)$-watermarked.
We suppose that $\rho_P>\rho_N$.
Simple Chernoff bounds on binomial variables show that, $\forall \tau\in(\rho_N,\rho_P)$:
\def\P{\mathbb{P}}
\def\Pfp{P_{\mathsf{fp}}}
\def\Pfn{P_{\mathsf{fn}}}
\begin{eqnarray}
\text{False positive}: \P(\rho_{n_d} > \tau) &\leq& e^{-n_d (\tau-\rho_N)^2},\\
\text{False negative}: \P(\rho_{n_d} < \tau) &\leq& e^{-n_d (\tau-\rho_P)^2}.
\end{eqnarray}
Given two maximum levels of errors $\Pfp$ (false positive) and $\Pfn$ (false negative), there exists a threshold $\tau\in(\rho_N,\rho_P)$ satisfying these requirements if
\begin{equation}
    n_d \geq \left(\frac{\sqrt{-\log \Pfp} + \sqrt{-\log \Pfn}}{\rho_P - \rho_N}\right)^2.
\end{equation}
We consider the worst hypotheses:
In the negative cas, the Verifier faces a Usurper who created a fake key. The black box is thus not watermarked with this fake key. The next section measures $\rho_N\approx 0.5$ at most. In the positive case, the black box is a modified version of the watermarked model. The next sections measures $\rho_P\approx 0.8$. Say $\Pfp=\Pfn=0.05$, these probabilities of errors are met for $n_d\approx 130$. 

\section{Experimental Results}\label{sec:experimentsResults}
\subsection{Setup}\label{sec:experimentsSetup}
The evaluation of the method uses MNIST and CIFAR10 datasets plus ImageNet for transfer learning, and off-the-shelf CNN network architectures (see App.~\ref{sec:App}).
We divide the training dataset into three parts: 80\% for training, 10\% for validation, and 10\% for fine-tuning in all of the experiments.

As for the attacks, we consider modifications of the model (like pruning, weight quantization, and fine-tuning) and modifications of the inputs before submitting to the model (like JPEG compression). We prune the DNN weights with rate $k\in(0,1)$ by setting randomly selected weights value to zero.
Weight quantization are done in four different ways. The weights are quantized into integers in dynamic range quantization (Dyn. Quant.) or in full unsigned 8 bits integer quantization (Full Uint8. Quant.), in full signed 8 bits integer quantization (Full Int8. Quant.), or converted to Float16 format (Float16 Quant.).
These operations reduce the size of the DNN and speed up querying time.
Fine-tuning uses the same algorithm than training (see Sect.~\ref{sec:App}) but with a learning rate of $10^{-5}$, for 30 epochs and a batch size of 64.

JPEG compresses and decompresses the input image before forwarding it to the CNN. We consider a low JPEG quality factor of 55.

\def\TA{\mathsf{TA}}
\def\Rectr{\mathsf{Rec_{tr}}}
\def\Rects{\mathsf{Rec_{ts}}}
\def\ASR{\mathsf{USR}}
The performance of the proposed method is evaluated with four criteria: $\TA$, $\Rectr$, $\Rects$, and $\ASR$.
$\TA$ represents the test accuracy of the model on the original task, $\Rectr$ is the probability associated to~\eqref{eq:rho} over 1,000 training mixup samples of $S_{tr}$, and $\Rects$ over 1,000 new mixup samples $S_{ts}$.
Finally, $\ASR$ is the usurper success rate, i.e. is the probability associated to~\eqref{eq:rho} measured on mixup samples generated with a fake key (picked by a Usurper). $\ASR$ is measured over 1,000 random fake keys.

For a given number of classes $m$, the selected classes to mixup and the mixup vector $(\lambdab,\mub)$ are all generated randomly.

As for the trigger $x_o$, it is an empty image with a white circle fixed at the top left.

As for the Usurper, in all the experiments, we grant him the knowledge of $m$, the statistical distribution of generating $\lambdab$ and $\mub$, and finally the overlay $x_o$. Granting the Usurper many secrets would validate the performance of the proposed scheme in extreme cases and would provide more emphasize to the Verifier about the value of the confidence of the proof. It is  difficult to know all the secrets in practice, unless for the case of an insider (the Usurper is the same as the DNN trainer or the outsource). 

\begin{table*}[htbp]
\caption{Performance Results of Mixer}
\label{tab:ResultsN2}
\centering
\begin{tabular}{|lcccccccc|}
\hline
\multicolumn{1}{|l||}{Metric}   & \multicolumn{1}{c||}{Host DNN} & \multicolumn{1}{l||}{Watermarked DNN} & \multicolumn{1}{l||}{Fine-Tune} & \multicolumn{1}{l||}{Dyn. Quant.} & \multicolumn{1}{l||}{Full Uint8. Quant.} & \multicolumn{1}{l||}{Full Int8. Quant.} & \multicolumn{1}{l||}{Float16 Quant.} & \multicolumn{1}{l|}{JPEG55} \\ \hline
\multicolumn{9}{|c|}{}                                                                                                                                                                                                                                                                            \\
\multicolumn{9}{|c|}{\textbf{MNIST}}                                                                                                                                                                                                                                                                                                           \\ \hline
\multicolumn{1}{|l||}{$\TA$}    & \multicolumn{1}{c||}{99.34}    & \multicolumn{1}{c||}{99.29}           & \multicolumn{1}{c||}{99.32}     & \multicolumn{1}{c||}{99.3}        & \multicolumn{1}{c||}{99.29}              & \multicolumn{1}{c||}{8.9}               & \multicolumn{1}{c||}{99.29}          & 99.12                       \\ \hline
\multicolumn{1}{|l||}{$\Rectr$} & \multicolumn{1}{c||}{-}        & \multicolumn{1}{c||}{100}             & \multicolumn{1}{c||}{100}       & \multicolumn{1}{c||}{100}         & \multicolumn{1}{c||}{100}                & \multicolumn{1}{c||}{0.0}                 & \multicolumn{1}{c||}{100}            & 100                         \\ \hline
\multicolumn{1}{|l||}{$\Rects$} & \multicolumn{1}{c||}{10.0}       & \multicolumn{1}{c||}{99.9}            & \multicolumn{1}{c||}{100}       & \multicolumn{1}{c||}{99.9}        & \multicolumn{1}{c||}{99.9}               & \multicolumn{1}{c||}{0.0}                 & \multicolumn{1}{c||}{99.9}           & 99.9                        \\ \hline
\multicolumn{9}{|c|}{}                                                                                                                                                                                                                                                                          \\
\multicolumn{9}{|c|}{\textbf{CIFAR10}}                                                                                                                                                                                                                                                                                                           \\ \hline
\multicolumn{1}{|l||}{$\TA$}    & \multicolumn{1}{c||}{83.99}    & \multicolumn{1}{c||}{84.69}           & \multicolumn{1}{c||}{84.59}     & \multicolumn{1}{c||}{84.57}       & \multicolumn{1}{c||}{84.51}              & \multicolumn{1}{c||}{9}                 & \multicolumn{1}{c||}{84.6}           & 77.01                       \\ \hline
\multicolumn{1}{|l||}{$\Rectr$} & \multicolumn{1}{c||}{-}        & \multicolumn{1}{c||}{100}             & \multicolumn{1}{c||}{100}       & \multicolumn{1}{c||}{100}         & \multicolumn{1}{c||}{100}                & \multicolumn{1}{c||}{0.0}                 & \multicolumn{1}{c||}{100}            & 90.8                        \\ \hline
\multicolumn{1}{|l||}{$\Rects$} & \multicolumn{1}{c||}{10.0}       & \multicolumn{1}{c||}{98.9}            & \multicolumn{1}{c||}{99.19}     & \multicolumn{1}{c||}{98.9}        & \multicolumn{1}{c||}{98.8}               & \multicolumn{1}{c||}{0.0}                 & \multicolumn{1}{c||}{98.9}           & 89.1                        \\ \hline
\multicolumn{9}{|c|}{}                                                                                                                                                                                                                                                                \\
\multicolumn{9}{|c|}{\textbf{Transfer Learning}}                                                                                                                                                                                                                                                                                                           \\ \hline
\multicolumn{1}{|l||}{$\TA$}    & \multicolumn{1}{c||}{86.54}    & \multicolumn{1}{c||}{86.07}           & \multicolumn{1}{c||}{85.5}      & \multicolumn{1}{c||}{86.0}          & \multicolumn{1}{c||}{85.9}               & \multicolumn{1}{c||}{9.1}               & \multicolumn{1}{c||}{86.07}          & 82.89                       \\ \hline
\multicolumn{1}{|l||}{$\Rectr$} & \multicolumn{1}{c||}{-}        & \multicolumn{1}{c||}{88.29}           & \multicolumn{1}{c||}{95.9}      & \multicolumn{1}{c||}{98.1}        & \multicolumn{1}{c||}{98.2}               & \multicolumn{1}{c||}{0.0}                 & \multicolumn{1}{c||}{98.3}           & 93.7                        \\ \hline
\multicolumn{1}{|l||}{$\Rects$} & \multicolumn{1}{c||}{10.0}       & \multicolumn{1}{c||}{88.59}           & \multicolumn{1}{c||}{84.6}      & \multicolumn{1}{c||}{88.6}        & \multicolumn{1}{c||}{88.6}               & \multicolumn{1}{c||}{0.0}                 & \multicolumn{1}{c||}{88.6}           & 82.7                        \\ \hline
\end{tabular}
\end{table*}

\subsection{Experimental Results}
\label{sec:results}

\subsubsection{Learning a secret manifold}
Table~\ref{tab:ResultsN2} first shows that the accuracy in recovering the watermark is very high either using training ($\Rectr$) or testing ($\Rects$) mixup samples. The lowest recovery accuracy is under transfer learning scenario with $\Rectr = 88.29\%$. Most importantly, the difference between $\Rectr$ and $\Rects$ is not statistically significant. It means that we succeed in making the model learn a secret manifold and not just a set of isolated points. This confirms that the Verifier no longer needs the trigger samples used for embedding the watermark. Also, from one verification to another, the queried samples differ.
As a scientific control, the probability~\eqref{eq:rho} is close to $1/C$ when verifying a vanilla unwatermarked model.

\subsubsection{Robustness against attacks}
\begin{figure}[!t]
\centerline{\includegraphics[width=0.99\columnwidth]{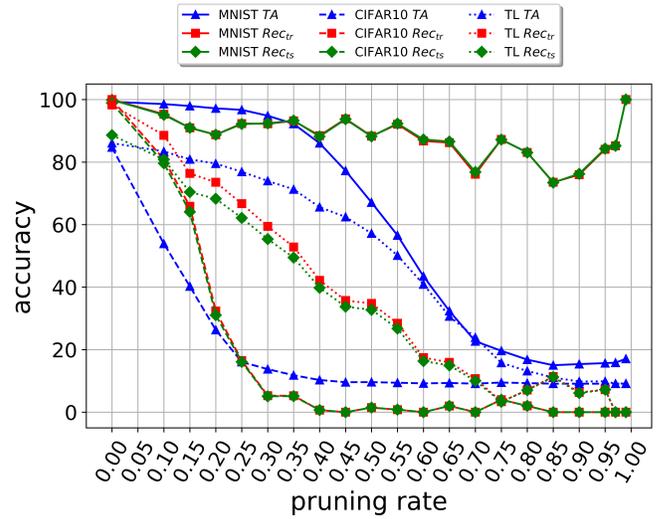}}
\caption{Accuracy vs. Pruning rate}
\label{fig:Pruning}
\end{figure}

From Table \ref{tab:ResultsN2}, we can observe that our watermarking scheme is robust to different kinds of post-processing that can be used to remove the watermark.

Starting by MNIST, by comparing the different metrics under attacks to the baseline, we observe that, under fine-tuning, the drop in $\TA$ is negligible and the recovery accuracies $\Rectr$, $\Rects$ remain almost intact. The same conclusion holds for Dynamic Quantization, full unsigned 8 bits integer quantization, Float16 quantization, and JPEG55. The worst attack is full signed 8 bits integer quantization which resets $\Rectr$ and $\Rects$ to absolute $0\%$. This huge drop in the the watermark recovery does not go without affecting the accuracy $\TA$ which is approximately to  random guess with a value of $8.9\%$ with MNIST, $9\%$ with CIFAR10, $9.1\%$ with transfer learning. This means that the attacker was not able to induce an error in the watermark recovery without ruining the main task the DNN is intended to perform, which simply makes the attacked DNN useless. 
JPEG compression with a quality factor of 55 has more severe effect on the performance. With the low jpeg quality factor considered, for MNIST case, $\TA$ drops by $0.17\%$, while $\Rectr$ and  $\Rects$ remains the same. The losses increases with CIFAR10, with a drop in $\TA$ drops by $7.68\%$, in $\Rectr$ by $9.2\%$, and in $\Rects$ by $9.8\%$. Even that CIFAR10 suffers more than MNIST from JPEG55, in the worst case, the watermark recovery remains above $89\%$. For transfer learning, $\TA$ drops by $3.18\%$, $\Rects$ by $5.89\%$, while $\Rectr$ increases by $5.41\%$. Therefore, also for JPEG55 compression, recovery rates does not drop without a heavy loss in the DNN main task accuracy.

Pruning the DNN has similar effect on the performance metrics compared to Table~\ref{tab:ResultsN2} attacks as illustrated in Figure~\ref{fig:Pruning}. Nonetheless, the drop in the accuracies is smoother for MNIST compared to CIFAR10 and transfer learning. MNIST $\TA$ resists above $60\%$ until $k=0.45$, while both $\Rectr$ and $\Rects$ remains above $70\%$. On the other hand, for CIFAR10, $\TA$, $\Rectr$ and $\Rects$ take a steep drop after $k=0.4$, while for transfer learning $k=0.15$ is sufficient to decrease all the metrics below $70\%$. Likewise, $\Rectr$ and $\Rects$ always follow any degradation in $\TA$.

As a conclusion, JPEG55,  Full Uint8. Quant. and pruning have more severe effect on the performance compared to other attacks. This is particularly correct for deeper networks with more complex tasks. Yet, this performance drop does not go without degrading the performance on the main task of the DNN which is, for the DNN owner is very beneficial as he can assure that his DNN will be only useful in the case where he can prove his ownership.
This intertwine between the performance losses is due to the fact that the mixup strategy embeds the watermark within the same features the DNN is intended to learn for its main task.

\subsubsection{Facing an Usurper}
The measured $\ASR$ are relatively high: $51.1\%$ for MNIST, $38.4\%$ for CIFAR10, and $39.5\%$ for transfer learning. 
This is due to the fact that, in the adopted scenario, we granted the attacker the knowledge of $m$, the statistical distribution for generating $\lambdab$ and $\mub$, and the overlay $x_o$.
We consider such an unrealistic threat model to show that, in the extreme case, it is difficult for the Usurper to have confidence as high as those reported in $\Rectr$ and $\Rects$ even under attack (provided the attacked model keeps a useful $\TA$).

\section{Conclusion and Future Works}\label{sec:conclusion}
This paper presents a new DNN watermarking protocol that uses image mixup for the construction and the injection of the watermark samples into the DNN.
The main properties of the proposed method is that it does not impair the DNN performance on the main task, it is robust against a wide range of attacks, and it creates a crucial inter-connection between the DNN main task and the watermarking task.
Future research directions are the security against more complex attacks, such as watermark overwriting, investigating the effect of the selection of the secrets in a predefined and precise manner. The theoretical investigation of the learning capacity of the network, and the input space dimension to understand the watermarking algorithm's limitations and capabilities would be another interesting direction.

\section{Acknowledgements}\label{sec:ack}
We would like to thank the ANR and AID french agencies for funding Chaire SAIDA ANR-20-CHIA-0011.

\section{Appendix}
\label{sec:App}
\subsubsection{MNIST}
The network for MNIST is as follows: 1 conv. layer (64 filters); a max pooling; 1 conv. (128 filters); a max pooling; 2 f.c. layers (256 and 10 neurons). For all the conv. layers, the kernel size is 5 with ReLU activation. The network is trained for 100 epochs with a batch size of 64. 

\subsubsection{CIFAR10 CNN}
The network for CIFAR10 is structured as follows: 2 conv. layers (32 filters); 2 conv. layers (64 filters); 2 conv. layers (128 filters); 2 conv. layers (256 filters) (each block of two conv. layers is followed by a $2 \times 2$ max-pooling and a dropout layer of rate 0.2); two f.c. layers (128 and 256 neurons) separated by a 0.2 dropout; final layer (10 neurons) with softmax activation. For all the conv. layers, the kernel size is $3$ with ReLU activation, initialized using He Uniform. The network is trained for 200 epochs with a batch size of 64.

\subsubsection{Transfer learning with ImageNet}
VGG19 pre-trained model on ImageNet is the base model for transfer learning. The network's decision part is replaced by two ReLU f.c. layers (1024 and 512 neurons) and the final layer (10 neurons) with softmax activation. Transfer learning is performed over CIFAR10 for 200 epochs with a batch size of 64.

During training, the cross entropy loss function is used in all cases. MNIST uses the Adam algorithm with default parameters, while CIFAR10 and transfer learning use SGD with learning rate 0.001 and momentum 0.9. 

\bibliographystyle{ieeetr}
\bibliography{mixup.bib}
\end{document}